\documentclass[english,pra,a4paper,twocolumn,preprintnumbers,showpacsnumber,superscriptaddress,amsmath,amssymb,notitlepage]{revtex4}
\usepackage{amsmath,amsfonts,amsthm,amssymb}
\usepackage{bbm}
\theoremstyle{plain}
\usepackage{graphicx}
\usepackage{dcolumn}
\usepackage{color}
\usepackage[colorlinks,linkcolor=blue, urlcolor=blue, anchorcolor=blue, citecolor=blue]{hyperref}
\usepackage{multirow}

\usepackage{babel}

\begin{document}
\title{Measure of quantum Fisher information flow in multi-parameter scenario}

\author{Haijun Xing}

\email{hjxing3@icloud.com}

\affiliation{Graduate School of China Academy of Engineering Physics, Beijing, 100193, China}

\author{Libin Fu}

\email{lbfu@gscaep.ac.cn}

\affiliation{Graduate School of China Academy of Engineering Physics, Beijing, 100193, China}

\begin{abstract}
We generalize the quantum Fisher information flow proposed by Lu \textit{et al}. [Phys. Rev. A \textbf{82}, 042103 (2010)] to the multi-parameter scenario from the information geometry perspective. A measure named the \textit{intrinsic density flow} (IDF) is defined with the time-variation of the intrinsic density of quantum states (IDQS). IDQS measures the local distinguishability of quantum states in state manifolds. The validity of IDF is clarified with its vanishing under the parameter-independent unitary evolution and outward-flow (negativity) under the completely positive-divisible map. The temporary backflow (positivity) of IDF is thus an essential signature of the non-Markovian dynamics. Specific for the time-local master equation, the IDF decomposes according to the channels, and the positive decay rate indicates the inwards-flow of the sub-IDF.
As time-dependent scalar fields equipped on the state space, the distribution of IDQS and IDF comprehensively illustrates the distortion of state space induced by its environment. As example, a typical qubit model is given.  
\end{abstract}

\maketitle

\section{introduction}

Information interchange between the open system and its environment is a critical viewpoint in studying the dynamics of open quantum systems.  Memory effect, i.e., the temporary revival of previously leaked information, is one of the fascinating topics in this fields \cite{Breuer2002}. It is considered a signature of the non-Markovianity firstly \cite{Rivas2014,Breuer2016,Vega2017,Li2019a,Li2019b,Breuer2009,Luo2012b,Lu2010,Laine2010}, then  became a vital method for manipulating quantum resources \cite{Breuer2009,Laine2010} such as entanglement \cite{Maniscalco2008,Bellomo2007,Mazzola2009,Huelga2012,Mirkin2019a,Mirkin2019b}, quantum interferometric power \cite{Dhar2015}, temporal steering \cite{Chen2016}, quantum coherence, correlations \cite{Luo2012b}, and quantum Fisher information (QFI) \cite{Lu2010,Song2015}. 
With the booming of technologies of control and manipulation open quantum systems \cite{Haikka2011,Yuan2017,Myatt2000,Liu2011,Gessner2014,Haase2018,Anderssen2019,Wu2020,Lu2020}, applications in ultracold atomic gases \cite{Haikka2011,Yuan2017}, quantum speed limit \cite{Deffner2013,Mirkin2016,Cimmarusti2015}, algorithms \cite{Dong2018,Roy2019}, and thermal machines \cite{Thomas2018,Abiuso2019} are under intensive  studies in recent years.

Specifically, in the conventional memory-free dynamics, the information leaks outwards from the system to the environment continuously, then dissolved. If the environment has nontrivial structures, the information may be memorized by the environment, then partially sending back into the system subsequently. Though the rigorous quantification of these memory effects highly depends on the interpretation of ``information," the distinguishability of quantum states is one of the primary choices. It can be captured by the trace distance \cite{Breuer2009,Laine2010} that measures the distinguishability of a pair of states, and quantum Fisher information (QFI) \cite{Lu2010,Song2015} that we focus on in this manuscript.

The QFI is an intrinsic measure of the local distinguishability of quantum states via estimating a given parameter. It has tight connections with the ``distance" measures between quantum states, such as the Fubini-Study metric \cite{Gibbons1992}, quantum geometric tensor \cite{Berry1989}, Bures distance \cite{Braunstein1994}, quantum fidelity \cite{Braunstein1994,Wootters1981}, and relative entropy \cite{Bengtsson2006}. Lu \textit{et al.} \cite{Lu2010} define the quantum Fisher information flow (QFIF) as a measure of the memory effect with the time-variation of QFI with respect to a parameter previously encoded into the probe state. It decomposes to sub-QFIF according to the channels of the time local master equation. The temporal appearance of inwards (positive) sub-QFIF is identified with the positive decay rates, hence becoming a signature of non-Markovian dynamics.  It performs well in single parameter cases. 
Furthermore, its applications in quantum metrology and quantum speed limits are fruitful: lots of achievements, both theoretical and experimental, have been made.

However, the practical systems are intrinsically multiple dimensional: 1) its states generally locate in multi-dimensional state space and thus is characterized by more than one parameter; 2) its dynamical evolution typically involves the variation of more than one parameter. The single-parameter scheme is therefore inadequate in thoroughly characterizing the dynamics of open systems. The generalization of QFIF to multi-parameter cases is essential.

In this article, we generalize the QFIF to the multi-parameter scenario from the information geometry perspective. We propose a measure named the \textit{intrinsic density flow} (IDF) with the time-variation of the intrinsic density of quantum states (IDQS), a fundamental information-theoretic state-distinguishability measure. The validity of IDF will be shown with its vanishing under the parameter-independent unitary evolution and negativity (outwards-flow) under the CP-divisible map. It indicates the positive (inwards) IDF is an essential signature of non-Markovian dynamics. Specific to the dynamics generated by the time-local master equation, the IDF is decomposable according to the channels. The direction (sign) of the sub-flow is determined by the decay rates: the temporary appearance of the positive decay rates indicates the backward sub-flow in the corresponding channel. It violates the CP-divisible condition and serves as a sufficient signature of the non-Markovian dynamics.

Furthermore, as time-dependent scalar fields equipped on the state space, the distribution of IDF and IDQS are potent tools to exhibit the detailed picture of the state space's  distorsion under the open dynamics. We will exemplify it with the typical model of a two-level system under the non-Markovian dissipative channels.  
  
This manuscript is arranged as follows: In Sec.~\ref{Review}, we review the QFI and QFIF from the information geometry perspective. In Sec.~\ref{IDQSIDF}, the IDF is introduced together with the IDQS. Specifically, a form of IDQS in the time-dependent coordinates is given in Sec.~\ref{EM}, for its tight connections with the equation of motion. In Sec.~\ref{FITLM}, the dynamics generated by the time local master equation are studied with the IDF. In Sec.~\ref{TLS}, the time-variation of IDQS and IDF are studied with a two-level system under the dissipative channels. At last, we conclude in Sec.~\ref{Conc}.   

\section{\label{Review}Review of QFI and QFIF from information geometry perspective}

In the formation geometry, a given state $\hat{\rho}(\boldsymbol{x}_0)$ is equivalent to a point $\boldsymbol{x}_0=(x^1_0,x^2_0,\dots,x^d_0)$ in the $d$-dimensional parameter space $\Omega_0$ via the model $\mathcal{S}_0=\{\hat{\rho}(\boldsymbol{x}_0)|\boldsymbol{x}_0\in \Omega_0\}$. A Riemannian metric $\boldsymbol{g}^F$ named as the quantum Fisher metric (QFM) with $d\times d$ entries
\begin{align}
 g^F_{\mu\nu}(\boldsymbol{x}_0)=&\frac{1}{8}\mathrm{Tr}\left[\{\hat{L}_\mu,\hat{L}_\nu\}\hat{\rho}\right],\label{QFM} 
\end{align}
$1\leqslant \mu,\nu\leqslant d$,
is equipped on the space $\Omega_0$, where $\{\cdot,\cdot\}$ is the anti-commutator, and the symmetric logarithmic derivative $\hat{L}_\mu$ is defined via
$
 \partial_\mu \hat{\rho}(\boldsymbol{x}_0)\equiv\frac{1}{2}\{\hat{\rho},\hat{L}_\mu \}
$
implicitly \cite{Helstrom1976,Holevo1982}, with $\partial_\mu\equiv \partial/\partial x_0^\mu$. The QFI $\boldsymbol{\mathcal{F}}$ is four times of the QFM $\boldsymbol{g}^F$. 
 
In the single-parameter metrology, the probe state $\hat{\rho}(\theta)$ sketches a curve $\boldsymbol{x}_0(\theta)$ in the initial parameter space $\Omega_0$  with the shift of parameter $\theta$. The estimation of $\theta$ is thus equivalent to identifying a point $\boldsymbol{x}_0(\theta)$ on the curve. The number of states distinguishable in a segment of the  curve $\boldsymbol{x}_0(\theta)$ is measured by the segment's length. The length is acquired by integrating the line element $ds$ with \cite{Wootters1981} 
\begin{equation} 
ds^2=\sum\nolimits_{\mu\nu}g^F_{\mu\nu}\dot{x}_0^\mu\dot{x}_0^\nu d\theta^2= g^F(\theta)d\theta^2,
\end{equation}
where $\dot{\boldsymbol{x}}_0\equiv d\boldsymbol{x}_0(\theta)/d\theta$ is the derivative of the curve $\boldsymbol{x}_0(\theta)$. Hence ${g}^F(\theta)^{1/2}$ is an intrinsic measure of the local density of states distinguishable on the curve $\boldsymbol{x}_0(\theta)$.
As the square of this density, the QFM itself is also a measure of the  distinguishability of $\hat{\rho}(\theta)$ from its neighboring quantum states. Furthermore, it is directly applicable in the quantum metrology as the upper bound of the estimator's precision \cite{GLM2004,GLM2006,GLM2011}.

When the dynamics $\Phi_t$ is applied, the states corresponding to the initial coordinates $\boldsymbol{x}_0$ is changed to $\hat{\rho}(\boldsymbol{x}_0;t)\equiv\Phi_t\hat{\rho}(\boldsymbol{x_0})$, with $\hat{\rho}(\boldsymbol{x}_0;0)\equiv\rho(\boldsymbol{x}_0)$. It forms a new state space $\mathcal{S}_t=\{\hat{\rho}(\boldsymbol{x}_0;t)=\Phi_t\hat{\rho}(\boldsymbol{x}_0)|\boldsymbol{x}_0\in \Omega_0\}$ at the given time $t$. Actually, $(\boldsymbol{x}_0;t)$ is a coordinates of the $(d+1)$-dimensional manifolds $\bigcup_t\mathcal{S}_t$ with $t$ as an additional dimension.

Under the dynamics $\Phi_t$, we also have $\hat{\rho}(\theta)\rightarrow\hat{\rho}(\theta;t)$. The corresponding QFI $\mathcal{F}(\theta;t)$ is lost (revival) with the variation of the state space $\mathcal{S}_t$. It can be accounted as the effects of outflow (inflow) of the information. Thus, Lu \textit{et al.} define the QFIF as \cite{Lu2010}
\begin{equation}
 \mathcal{I}_{\mathrm{LWS}}(\theta;t)\equiv\frac{d\mathcal{F}(\theta;t)}{dt}=4\frac{dg^F(\theta;t)}{dt},
\end{equation}
i.e., the time variation of the QFI. From the information geometry perspective,
it measures the time-variation of the square of the density of distinguishable states along the given curve $\boldsymbol{x}_0(\theta)$.

\section{\label{IDQSIDF}Intrinsic density of quantum states and Intrinsic density flow}

\subsection{Intrinsic density of quantum states}

To identify a state $\hat{\rho}(\boldsymbol{x}_0;t)\in \mathcal{S}_t$ in general cases, one should acknowledge all of the components of $\boldsymbol{x}_0$ at a given time $t$. It is equivalent to localizing the point $\boldsymbol{x}_0$ in the $d$-dimensional initial parameter space $\Omega_0$, where $\hat{\rho}(\boldsymbol{x}_0;t)$ is surrounded by states in all of the ``directions." Furthermore, the dynamical evolutions generally affect the distinguishability of quantum states in the multi-direction of the state space $\mathcal{S}_t$. A single element of the QFM, i.e., $g^F(\theta)$, is thus inadequate for characterizing the distinguishability of $\hat{\rho}(\boldsymbol{x}_0;t)$ out of its neighborhood.    

Theoretically, the local statistical distinguishability of quantum states in $\Omega_0$ can be measured with the intrinsic density of quantum states (IDQS) \cite{Xing2020}
\begin{equation}
	\mathcal{D}_Q(\boldsymbol{x}_0;t)=\sqrt{|\boldsymbol{g}^F(\boldsymbol{x}_0;t)|}=\frac{dV(\mathcal{S}_t)}{d^d\boldsymbol{x}_0},
\end{equation}
with $|\boldsymbol{A}|$ denoting the determinant of matrix $\boldsymbol{A}$, where the invariant volume element  $dV(\mathcal{S}_t)$ quantifies the number of quantum states locating in the element $d^d\boldsymbol{x}_0$. 
We mention that for the pure state, IDQS is the measure that defines the completeness relationship of (sub-manifolds of) the projective Hilbert spaces \cite{Xing2020,Bengtsson2006}.

\subsection{Intrinsic density flow}
Although the state $\hat{\rho}(\boldsymbol{x}_0;t)$ is evolving under the map $\Phi_{t}$, the corresponding point $\boldsymbol{x}_0$ is stationary in the initial parameter space $\Omega_0$. On the contrary, the QFM, thereby IDQS, is time-dependent and capable of characterizing the distortion of state space under the map. Specifically, IDQS is a qualified ``information" measure that meets the essential criteria \cite{Ruskai1994,Breuer2009,Breuer2016} satisfied by the trace distance: IDQS is non-negative, invariant for parameter-independent unitary dynamics, and contraction under the parameter-independent completely positive and trace-preserving (CPTP) maps. In a concise form,  we have (for proof, see Appendix  \ref{Vanishes}, \ref{contraction}, and \ref{SLD})
\begin{equation}
 \mathcal{D}_Q(\boldsymbol{x}_0;t)\geqslant\mathcal{D}_Q(\boldsymbol{x}_0;t')\geqslant0,\label{DCPTP}
\end{equation}
for arbitrary given state $\hat{\rho}(\boldsymbol{x}_0;t)$ and $\hat{\rho}(\boldsymbol{x}_0;t')\equiv\Lambda_{t',t}[\hat{\rho}(\boldsymbol{x}_0;t)]$ with $\Lambda_{t',t}$ denoting an arbitrary parameter-independent CPTP map,
where the first equality is reached by the unitary channel 
$\Lambda_{t',t}[\cdot]=\hat{U}(t,t')\cdot \hat{U}^\dagger(t,t')$ with $\partial_\mu \hat{U}(t,t')=0$.  The unitary invariance of $\mathcal{D}(\boldsymbol{x}_0;t)$ indicates IDQS measuring an information which conservative in the composite of system and environment. The contraction of IDQS indicates the revival information is always smaller than the previous leaking information.
      
Based on the above discussions, we define the \textit{intrinsic density flow} (IDF) as
\begin{equation}
 \mathcal{I}(\boldsymbol{x}_0;t)\equiv\frac{d}{d t}\mathcal{D}_Q(\boldsymbol{x}_0;t),
\end{equation}
i.e., the time-variation of the IDQS.
Its negative value indicates the leaking of information from system to environment.  The positive value indicates the backflow of the previous leaking information, which is a signature of the non-Markovianity. Specifically, the IDF has the following properties:
\begin{description}
\item[{a1}] IDF is not positive under the parameter-independent CP-divisible dynamics and vanishes under the parameter-independent unitary dynamics. It directly results from Eq.~(\ref{DCPTP}) and makes $\mathcal{I}(\boldsymbol{x}_0;t)>0$ a sufficient condition for the non-CP divisible dynamics.
\item[{a2}] IDF is a linear function of $d\hat{\rho}/dt$ and $\mathcal{D}(\boldsymbol{x}_0;t)$ as (for details, see Appendix~\ref{SLD})
\begin{equation}
\mathcal{I}(\boldsymbol{x}_0;t)=\mathcal{I}^R(\boldsymbol{x}_0;t)\mathcal{D}(\boldsymbol{x}_0;t),
\end{equation}
with the \emph{relative intrinsic density flow} (RIDF)
\begin{equation}
\mathcal{I}^R(\boldsymbol{x}_0;t)\equiv\frac{1}{2}\mathrm{tr}\left[\frac{1}{\boldsymbol{g}^F}\mathrm{Tr}(\boldsymbol{\mathcal{L}}\frac{d\hat{\rho}}{dt})\right],
\end{equation}
where $\mathrm{Tr}$ (tr) denotes the trace operation in Hilbert space (parameter space), $\boldsymbol{\mathcal{L}}$ is a $d\times d$-dimensional matrix with the entry
\begin{equation}
 \hat{\mathcal{L}}_{\mu\nu}=\frac{1}{2}\left[\hat{L}_\nu(2\hat{\partial}_\mu-\hat{L}_\mu)+\hat{L}_\mu (2\hat{\partial}_\nu-\hat{L}_\nu)\right].
\end{equation}
IDF thus inherits the linear structures of the master equation, as shown in Sec. \ref{FITLM}.   
\item[{a3}] By choosing $\boldsymbol{x}_0$ as a complete basis, i.e., a coordinate system of the state space concerned, the IDQS $\mathcal{D}(\boldsymbol{x}_0;0)$ captures all the local information of the initial state $\rho(\boldsymbol{x}_0;0)$. Then effects of the dynamical evolution $\Phi_{t}$ on all of the components of $\boldsymbol{x}_0$ are accounted in $\mathcal{I}(\boldsymbol{x}_0;t)$. 
\item[{a4}] The RIDF is independent of the parameterization model. For two time-independent 
coordinates $\boldsymbol{x}_0$ and $\boldsymbol{y}_0$ of the initial parameter space $\Omega_0$, we have
\begin{equation}
 \mathcal{I}^R(\boldsymbol{x}_0;t)=\mathcal{I}^R(\boldsymbol{y}_0;t).\label{ChangeCo}
\end{equation}
The corresponding IDFs only differ in a time-independent constant as 
$
\mathcal{I}(\boldsymbol{x}_0;t)=\left|\frac{\partial\boldsymbol{y}_0}{\partial\boldsymbol{x}_0}\right| \mathcal{I}(\boldsymbol{y}_0;t)
$. 
Hence, the direction of IDF is also independent of the parameterization model.
\end{description}
These properties make the IDF be a qualified measure of the local information flow. Before further studies, we will introduce another form of IDF.

\subsection{\label{EM}State space $\mathcal{S}$ and IDF}

In practical studies, researchers favor to identify the state space $\mathcal{S}_t$ at different times with $(\boldsymbol{x}_0;t)\sim(\boldsymbol{x}'_0;t')$ if $\hat{\rho}(\boldsymbol{x}_0;t)=\hat{\rho}(\boldsymbol{x}'_0,t')$. We denote it as the state space $\mathcal{S}\equiv\{\hat{\rho}(\boldsymbol{x})=\hat{\rho}(\boldsymbol{x}_0;t)|\boldsymbol{x}_0\in\Omega_0,t\}$, where $\boldsymbol{x}$ serves as the coordinates of the state space $\mathcal{S}$. A dynamical evolution is thus depicted by the movement in $\mathcal{S}$: the initial state $\hat{\rho}(\boldsymbol{x}_0)$ sketches an orbit illustrated by the equations of motion $\boldsymbol{x}(t)$ with $\boldsymbol{x}(0)=\boldsymbol{x}_0$. One can define an alternative IDQS in space $\mathcal{S}$ as $\mathcal{D}_Q(\boldsymbol{x})=\sqrt{|\boldsymbol{g}^F(\boldsymbol{x})|}$ with respect to the coordinates $\boldsymbol{x}$.  

\begin{figure}
\centering
\includegraphics[width=5cm]{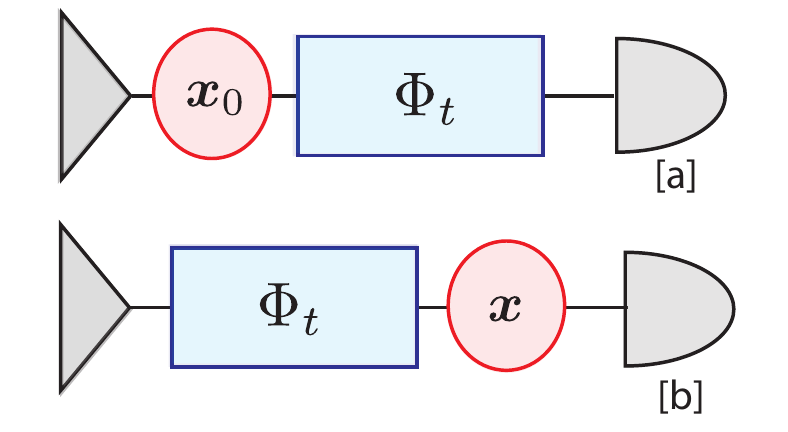}
\caption{(color online). Two coordinates (parameterization schemes) form the quantum metrology perspective. (a) The initial parameter $\boldsymbol{x}_0$ are encoded into the probe $\rho$ firstly, then sent to the channel $\Phi_{t}$. The precision of $\boldsymbol{x}_0$'s estimation is upper bounded by the QFM $\boldsymbol{g}^F(\boldsymbol{x}_0;t)$. (b) The  parameter $\boldsymbol{x}$ are encoded after the state through the channel $\Phi_{t}$. The precision of $\boldsymbol{x}$'s estimation is upper bounded by the QFM $\boldsymbol{g}^F(\boldsymbol{x})$.}  
\label{Fig2}
\end{figure}     

For state $\hat{\rho}(\boldsymbol{x})$ with the equation of motion $\boldsymbol{x}(t)$, we have the IDQS
$
\mathcal{D}_Q(\boldsymbol{x}_0;t)=\left|\frac{\partial\boldsymbol{x}}{\partial\boldsymbol{x}_0}\right|\mathcal{D}_Q(\boldsymbol{x}), 
$
and the corresponding IDF decomposes as
\begin{align}
    \mathcal{I}(&\boldsymbol{x}_0;t)
 =  \mathcal{D}_Q(\boldsymbol{x}_0;t)\nonumber\\
   &\times \left\{\frac{1}{2}\mathrm{tr}\left[\frac{d}{dt}\log\boldsymbol{g}^F(\boldsymbol{x})\right]+\mathrm{tr}\left[\frac{d}{dt}\log\begin{pmatrix}\frac{\partial\boldsymbol{x}}{\partial\boldsymbol{x}_0}\end{pmatrix}\right]\right\},\label{EMD}
\end{align}
where the first term describes shift of the point $\boldsymbol{x}(t)$ along its orbit in the state space $\mathcal{S}$. The second term is contributed by the Jacobian which connect the state space $\mathcal{S}$ and initial parameter space $\Omega_0$ . From the initial information perspective, it depicts the variation of the frame $\boldsymbol{x}$, i.e., the \emph{background geometry} of $\mathcal{S}$. 

We mention that, from the quantum metrology perspective, the coordinates $\boldsymbol{x}_0$ and $\boldsymbol{x}$ are parameters encoded into the probe and awaiting estimation. As shown in Fig.~\ref{Fig2}a (b), $\boldsymbol{g}^F(\boldsymbol{x}_0;t)$ ($\boldsymbol{g}^F(\boldsymbol{x})$) depicts the QFI acquired via parameterizing states before (after) the dynamical evolution $\Phi_{t}$. In case Fig.~\ref{Fig2} b, the channel $\Phi_{t}$ is actually part of the state preparation. 

\section{\label{FITLM}Intrinsic density flow with time-local master equation}

In this section, we study the dynamics of open quantum systems with the IDF. Specifically, we focus on the state $\hat{\rho}(\boldsymbol{x}_0;t)$ whose evolution is governed by the time-local master equation \cite{Gorini1976,Lindblad1976,Breuer2004,Breuer2016}
\begin{equation}
 \frac{d}{dt}\hat{\rho}(\boldsymbol{x}_0;t)=\mathcal{K}(t)\hat{\rho}(\boldsymbol{x}_0;t),
\end{equation}
with the generator $\mathcal{K}(t)$ acting on the state $\hat{\rho}$ as 
\begin{equation}
 \mathcal{K}(t)\hat{\rho}=-\mathrm{i}[\hat{H},\hat{\rho}]+\sum_{i}\gamma_{i}\left[\hat{A}_{i}\hat{\rho} \hat{A}_{i}^{\dagger}-\frac{1}{2}\{\hat{A}_{i}^{\dagger}\hat{A}_{i},\hat{\rho}\}\right],\label{TLME}
\end{equation}
where all of the $\hat{H}$, $\gamma_i$, and $\hat{A}_i$ are generally time-dependent. It is a generalization of the conventional Lindblad master equation that all $\hat{A}_i$ and $\gamma_i$ are time-independent, and $\gamma_i$ are non-negative. Eq.~(\ref{TLME}) leads a CP-divisible dynamic, if and only if $\gamma_i$ is non-negative for all channel $\hat{A}_i$ at all of the time \cite{Gorini1976,Lindblad1976,Breuer2016}. Hence, the temporary appearance of negative $\gamma_i$ is taken as the signature of the non-Markovian (non CP-divisible) dynamics. It is also necessary for the memory effects and backflow of the information \cite{Piilo2008,Breuer2009b}. We further assume 
$
 \partial_\mu\hat{H}=0$ and $\partial_\mu\hat{A}_i=0.
$ It indicates the linearity of the von Neumann equation and inconsistent with the quantum no-cloning theorem \cite{Lu2010,Wootters1982}. 
It also makes state $\hat{\rho}(\boldsymbol{x}_0;t)$ a stationary point in the initial parameter space.

Firstly, we focus on a special case with $\gamma_i=0$ $\forall i$, where Eq.~(\ref{TLME}) reduces to the unitary evolution. The corresponding IDF vanishes as
$
\mathcal{I}(\boldsymbol{x}_0;t)_0=0\label{IU}
$, with the additional footnote $0$ denoting the unitarity.
It directly results from the time-invariance of the metric under unitary dynamics with (for proof, see Appendix \ref{Vanishes})
\begin{align}
 \frac{d}{d t} g^F_{\mu\nu}(\boldsymbol{x}_0;t)_0 = 0.\label{LiouE}
\end{align}
It indicates that the parameter space $\Omega_0$ is frozen. It is tremendously different from the picture in state space $\mathcal{S}$. In the coordinates $\boldsymbol{x}$, the system may demonstrate very complicated dynamics.

In the general cases of Eq.~(\ref{TLME}), the system exchanges information with the environment through each of the channel $\hat{A}_i$ with $\gamma_i\neq 0$. Specifically, the IDF decomposes as $\mathcal{I}(\boldsymbol{x}_0;t)
 =  \sum_i\mathcal{I}_i(\boldsymbol{x}_0;t)$, where 
\begin{equation}
 \mathcal{I}_i(\boldsymbol{x}_0;t) \equiv   
\frac{\gamma_i}{2}\mathrm{tr}\left[\frac{1}{\boldsymbol{g}^F}\left( \frac{d}{d t}\boldsymbol{g}^F\right)_i\right] \mathcal{D}_Q(\boldsymbol{x}_0;t)\label{Sub-IDF}
\end{equation}     
denotes the sub-IDF through the channel $\hat{A}_i$ with the derivatives (for details, see Appendix~\ref{FTLME})
\begin{equation}
 (\frac{d}{dt}g^F_{\mu\nu})_i
  = \frac{1}{2}\mathrm{tr}\left\{([\hat{A}_i^{\dagger},\hat{L}_\nu][\hat{A}_i,\hat{L}_\mu ]+[\hat{A}_i^{\dagger},\hat{L}_\mu][\hat{A}_i,\hat{L}_\nu ])\hat{\rho}\right\}. 
\end{equation}
For the matrix $(d\boldsymbol{g}^F/dt)_i$ is negative semidefinite, we have the direction (sign) of sub-flows
\begin{equation}
 \mathcal{I}_i(\boldsymbol{x}_0;t)\begin{cases}
 \geqslant 0, & \gamma_i(t)<0\\
 = 0,		  & \gamma_i(t)=0\\
 \leqslant 0, & \gamma_i(t)>0.
\end{cases}\label{Subflow}
\end{equation}
 
These results are full of physical implications: (1) The decomposition of IDF according to the channel results from the time-local master equation's linearity to $d\hat{\rho}/dt$. (2) The direction of the sub-flow $\mathcal{I}_i$ is controlled by $\gamma_i$. If there exist a channel such that $\gamma_i<0$ for some time $t$, the corresponding sub-flow will flow back to the system. It is consistent with the CP-divisible condition given by the Gorimi-Kossakowski-Sudarshan-Lindblad theorem \cite{Gorini1976,Lindblad1976}. We mention that these results are  the natural generalization of the proposition Eq.~(6) in \cite{Lu2010}. However, it is now valid in the multi-parameter scenario and independent of the parameterization scheme.

\section{\label{TLS}Two-level systems}

For two-level system with basis $\{|0\rangle,|1\rangle\}$, we parameterize the general mixed state $\hat{\rho}$ as
\begin{equation}
 \hat{\rho}(\boldsymbol{n})=\frac{1}{2}\left(\hat{\mathbb{I}}_2+\boldsymbol{n}\cdot\hat{\boldsymbol{\sigma}}\right),
\end{equation}
 with the Pauli matrices $\hat{\boldsymbol{\sigma}}=(\hat{\sigma}_1,\hat{\sigma}_2,\hat{\sigma}_3)$, where $\mathcal{S}$ is the Bloch sphere, and coordinates $\boldsymbol{x}$ is the Bloch vector $\boldsymbol{n}=(n^1,n^2,n^3)$ , with $|\boldsymbol{n}|^2\equiv\sum_\mu (n^\mu)^2\leqslant 1$. We have the density
\begin{equation}
\mathcal{D}_Q(\boldsymbol{n})\equiv\sqrt{|\boldsymbol{g}^F(\boldsymbol{n})|}=\frac{1}{8\sqrt{1-|\boldsymbol{n}|^2}}.\label{DDSQBit}
\end{equation}
This density only depends on the radius $|\boldsymbol{n}|$, i.e., the state's  purity. It results from the unitary invariance of IDQS. 
$\mathcal{D}_Q(\boldsymbol{n})$ is divergent if $|\boldsymbol{n}|=1$. It is induced by the radial element $g^F_{nn}(\boldsymbol{n})=1/[4(1-|\boldsymbol{n}|^2)]$, which depicts the distinct statistical difference between pure and mixed states. However, the statistical distance between any pair of pure and mixed states are still finite, so is the volume given by this density. The minimum of $\mathcal{D}_Q(\boldsymbol{n})$ is reached by the completely mixed state with $\boldsymbol{n}=0$. Surprisingly, this minimum is non-zero. It depicts the statistical difference between $\hat{\rho}(\boldsymbol{0})$ and its neighboring states, although $\hat{\rho}(\boldsymbol{0})$ itself is usually termed as information-free.

\begin{figure*}
\centering
\includegraphics[width=17.0cm]{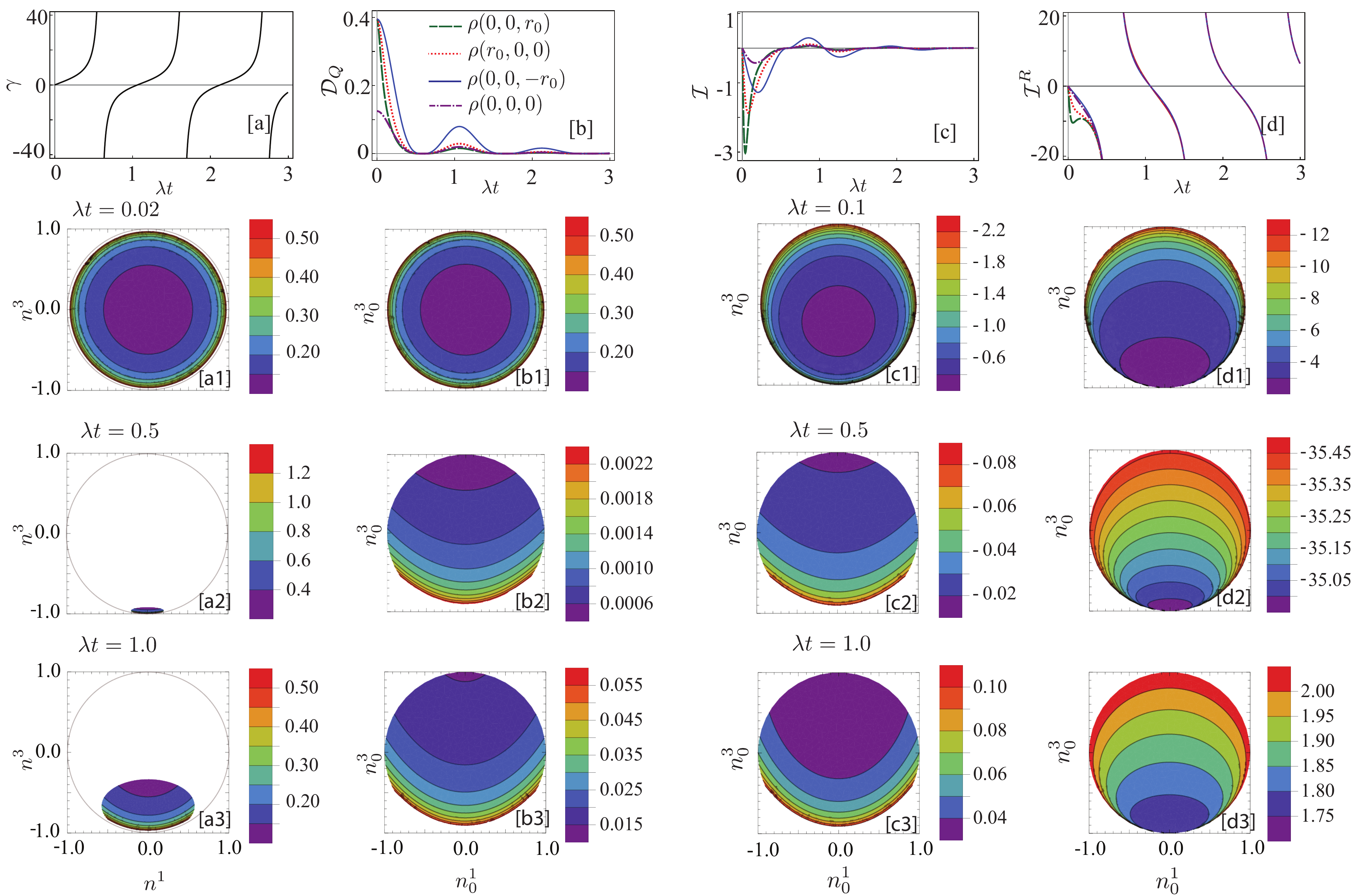}
\caption{(color online). Two-level systems in dissipative channel ($W=3\lambda$). (a) $\gamma$ as a function of time; (b) The IDQS $\mathcal{D}_Q(\boldsymbol{n}_0;t)$ as a function of time, with $r_0=\sqrt{0.9}$; (c) the IDF $\mathcal{I}(\boldsymbol{n}_0;t)$ as a function of time; (d) the RIDF $\mathcal{I}^R(\boldsymbol{n}_0;t)$ as a function of time. (a1)-(a3)  $\mathcal{D}_Q(\boldsymbol{n})$ in $n^1$-$n^3$ plane with $n^2=0$ at $\lambda t=0.02$, $0.5$, and $1.0$; (b1)-(b3) The IDQS $\mathcal{D}_Q(\boldsymbol{n}_0;t)$ in $n^1_0$-$n^3_0$ plane with $n_2^0=0$ at $\lambda t=0.02$, $0.5$, and $1.0$. (c1)-(c3)  $\mathcal{I}(\boldsymbol{n_0};t)$ in $n^1_0$-$n^3_0$ plane with $n^2_0=0$ at $\lambda t=0.1$, $0.5$, and $1.0$; (d1)-(d3) $\mathcal{I}^R(\boldsymbol{n_0};t)$ in $n^1_0$-$n^3_0$ plane with $n_2^0=0$ at $\lambda t=0.1$, $0.5$, and $1.0$.}  
\label{Fig3}
\end{figure*}

\subsection{Dissipative channels}

In this sub-section, we study a typical model where a two-level system is immersed  in a dissipative environment \cite{Breuer1999,Lu2010,Breuer2016}. In the interaction picture, the system undergoes a dynamic generated by the master equation  
\begin{align}
    \frac{d}{dt}\hat{\rho}(t)
 =& \gamma(t)\left[\hat{\sigma}_{-}\hat{\rho}\hat{\sigma}_{+}-\frac{1}{2}\{\hat{\sigma}_{+}\hat{\sigma}_{-},\hat{\rho}\}\right]\label{DispDyn},
\end{align}
with the raising (lowering) operator $\hat{\sigma}_{+}$ $(\hat{\sigma}_{-})$. By assuming the environment has a Lorentzian spectral density with vanishing detuning, we have $\gamma(t)=-2\dot{h}(t)/h(t)$ with the characteristic function
\begin{equation}
h(t)=\begin{cases}
 e^{-\lambda t/2}\left[\cosh(\frac{dt}{2})+\frac{\lambda}{2}\sinh(\frac{dt}{2})\right], & W<\frac{\lambda}{2}\\
e^{-\lambda t/2}\left[\cos(\frac{dt}{2})+\frac{\lambda}{2}\sin(\frac{dt}{2})\right],  & W\geqslant\frac{\lambda}{2},
\end{cases}
\end{equation}
$d=\sqrt{|\lambda^2-4W^2|}$, where $W$ measures the system-environment coupling strength, $\lambda$ defines the width of the spectral density. 
In the weak coupling regime with $W<\lambda/2$, $\gamma(t)$ is non-negative. The system undergoes a Markovian dynamics, where the information---measured by both of the trace  distance \cite{Laine2010} and QFI \cite{Lu2010}---is lost continuously. For conciseness, we mainly focus on the strong coupling regime with $W\geqslant\lambda/2$, where $h(t)$ displays an oscillatory behavior and the non-Markovian dynamics emergent with the negative $\gamma(t)$.

We begin with the equations of motion
\begin{subequations}
\begin{align}
 n^{\mu}(t) & =h(t)n^{\mu}_0, \quad \mu=1,2,\\
 n^{3}(t) & =h(t)^{2}(1+n^{3}_0)-1.
\end{align}
\end{subequations}
Under this equation, the Bloch sphere contracts to ground state $|0\rangle$ firstly with the positive $\gamma(t)$, then partially swell back corresponding to the negative $\gamma(t)$, as shown by Fig.~\ref{Fig3} a1-a3.
Together with Eq.~(\ref{EMD}) and (\ref{DDSQBit}), we have the IDF
\begin{equation}
 \mathcal{I}(\boldsymbol{n}_0;t)=-\frac{\gamma}{2}\left[3+\frac{(1+n^3)^2}{1-|\boldsymbol{n}|^2}\right]\mathcal{D}_Q(\boldsymbol{n}_0;t)
\end{equation}
with the IDQS
\begin{equation}
\mathcal{D}_Q(\boldsymbol{n}_0;t)=\frac{h(t)^4}{8\sqrt{1-|\boldsymbol{n}|^2}}.
\end{equation}
As shown by Fig.~\ref{Fig3}a, $\gamma(t)$ oscillates between the positive and negative values. It induces the information flow $\mathcal{I}(\boldsymbol{n}_0;t)$ to propagate outwards and inwards, and the IDQS $\mathcal{D}_Q(\boldsymbol{n}_0;t)$  increases and decreases correspondingly. The time-local extremum of the IDQS is given at the transition times of inwards and outwards flow, when the IDQS vanishes with the Bloch sphere shrinks to the point $|0\rangle$. 
In Fig.~\ref{Fig3}b (\ref{Fig3}c), we illustrate the IDQS (IDF) of four initial states: $\hat{\rho}_1(0,0,r_0)$, $\hat{\rho}_2(r_0,0,0)$, $\hat{\rho}_3(0,0,-r_0)$, and $\hat{\rho}_4(0,0,0)$, with blue, red, green, and purple lines, respectively. At $t=0$, $\hat{\rho}_1$, $\hat{\rho}_2$, and $\hat{\rho}_3$ locate in the same shell with radius $r_0=\sqrt{0.9}$, the maximally mixed state $\hat{\rho}_4$ locates in the center of the space. They show the same time-dependent non-Markovian behavior.
The state $\hat{\rho}_3$ near the stationary point $|0\rangle$ has the biggest IDQS over the four states. We mention that the distinguishability of the ``information free" state $\hat{\rho}_4$ from its neighboring states are still captured by IDQS and IDF, which exhibited the same dynamical characteristics as other states.
 
For both the IQDS and IDF are scalar fields equipped on the parameter space $\Omega_0$,  the variation of their distributions provides us an interesting viewpoint to study the dynamics of the open system. As shown by Fig. \ref{Fig3}~c2 and c3, the IDF takes its maximum at point $|1\rangle$ in the beginning of the first contraction. It induces the 
tremendous decrease of the IDQS for states in its neighborhood, as shown by Fig.~\ref{Fig3} b2 and exemplified by state $\hat{\rho}_1$.
The IDQS always takes its maximum (minimum) in the point $|0\rangle$ ($|1\rangle$) in the subsequent evolutions. It is induced by the fact that the inward flow's magnitude is always smaller than the previous outwards flow, although the flow takes its maximum at state $|0\rangle$ in the following stages.     

Furthermore, the RIDF captures a parameterization-independent signature of the dynamics. As a time-dependent scalar field equipped on the parameter space, its distribution is an overall description of the relative strength of state space's distortion. Specific for this dissipative model, we have
\begin{equation}
\mathcal{I}^R(\boldsymbol{n}_0;t)=-\frac{\gamma}{2}\left[3+\frac{(1+n^3)^2}{1-|\boldsymbol{n}|^2}\right].
\end{equation}
Its distribution is shown by Fig.~\ref{Fig3} d1-d3. It indicates the IDQS near point $|1\rangle$ is lost and acquired with a more significant relative strength, not only in the first contraction but also in the full evolutions. It consists of the insights that the excited states component $|1\rangle\langle1|$ is mostly influenced by the dissipative channel. Furthermore, the gradient and range of $\mathcal{I}^R(\boldsymbol{n}_0;t)$ indicate the uniformity of the time-variation of IDQS fields. As exemplified by Fig.~\ref{Fig3} d and d1, the IDQS leaks with a larger gradient in the first contraction. In the following stage, as shown by Fig.~\ref{Fig3} d2 and d3, the gradient and range are tremendously decreased. The IDQS at all of the points are lost and revival with roughly the same RIDF. It results from that the Bloch sphere shrinks to an ellipsoid highly localized around the point $|0\rangle$ after the first contraction. 

\section{\label{Conc}Conclusions}
In conclusion, we have generalized the quantum Fisher information flow (QFIF) \cite{Lu2010} to the multi-parameter scenario from the information geometry perspective. We propose a measure named the intrinsic density flow (IDF) with the time-variation of quantum states' local distinguishability, quantified by the intrinsic density of quantum states. The validity of IDF has been shown with its vanishing under the parameter-independent unitary evolution and negativity (propagate outwards) under the completely positive and trace-preserving map. It makes the positive (inwards) IDF be a vital signature of non-Markovian dynamics. 

Specific for dynamics generated by the time local master equation, we have shown the IDF is decomposable according to the channels. The direction (sign) of the sub-flow is determined by the decay rates: the temporary appearance of the positive decay rates indicates the sub-flow in the corresponding channel  flowing backward. It violates the CP-divisible condition and serves as an essential signature of the non-Markovian dynamics.

Not only having tight connections with the non-Markovian dynamics, but the IDQS and IDF themselves are also significant for the studies of open quantum systems.
As time-dependent scalar fields equipped on the state space, the distribution of IDF and IDQS are potent tools to exhibit the global picture of the state space's variation under the open dynamics. We have exemplified it with the qubit system under the non-Markovian dissipative channel.  

 Via ten years of productive studies, the QFIF exhibits its values in both theoretical and experimental aspects. Our research and the IDF provide a path to generalize these studies to the multi-parameter cases. Furthermore, its tight connections with the information geometry have been built, which gives us the systematic methods to study the dynamics of the open quantum systems. We mention that IDF is one of many measures provided by the quantum Fisher metric, though possibly the most important one. Other explorations are in processing. The potential value of this field is promising. We expect this article can catalyze more studies of the open quantum system from the information geometrical perspective.

\begin{acknowledgements}
This work is supported by the National Natural Science Foundation of China (NSFC) (Grant No. 12088101, No. 11725417, and No. U1930403) and Science Challenge Project (Grant No. TZ2018005). 
\end{acknowledgements}

\appendix

\section{\label{Vanishes}Proof of Eq.~(\ref{LiouE}): The IDF vanishes under the parameter-independent unitary channel.}
We will prove the invariance of QFM metric
\begin{equation}
 \frac{d}{dt}{\boldsymbol{g}}^F(\boldsymbol{x}_0;t)=0,\label{IUD}
\end{equation}
under the unitary evolution $U$ generated by
\begin{equation} 
\frac{d}{dt}\rho(\boldsymbol{x}_0;t)=-\mathrm{i}[H,\rho],
\end{equation}
where $H$ is assumed Hermitian and parameter-independent with $H=H^\dagger$ and $\partial_\mu H=0$. 
\begin{proof}
We denote the symmetric logarithmic derivative of $\rho(\boldsymbol{x}_0;0)$ as $L_\mu (\boldsymbol{x}_0;0)$, which satisfying
$
\partial_\mu\rho(\boldsymbol{x_0};0)=\frac{1}{2}\{\rho(\boldsymbol{x}_0;0),L(\boldsymbol{x}_0;0)\}$, and $L_\mu^\dagger(\boldsymbol{x}_0;0)=L_\mu(\boldsymbol{x}_0;0)$.
Then 
\begin{equation}
L_\mu(\boldsymbol{x}_0;t)= U(t,0)L(\boldsymbol{x}_0;0)U^\dagger(t,0) 
\end{equation}
is a valid symmetric logarithmic derivative of the state 
 $\rho(\boldsymbol{x}_0;t)=U(t,0)\rho(\boldsymbol{x}_0;0)U^\dagger(t,0)$, which satisfies $\partial_\mu\rho(\boldsymbol{x}_0;t)=\frac{1}{2}\{L_\mu(\boldsymbol{x}_0;t),\rho(\boldsymbol{x}_0;t)\}$ and $L^\dagger_\mu(\boldsymbol{x}_0;t)=L_\mu(\boldsymbol{x_0};t)$. Insert it into the definition of QFM Eq.~(\ref{QFM}), we have
\begin{eqnarray}
	 g^F_{\mu\nu}(\boldsymbol{x}_0;t)
 &=& \frac{1}{8}\mathrm{tr}\left[\left\{L_\mu(\boldsymbol{x}_0;t),L_\nu(\boldsymbol{x}_0;t)\right\}\rho(\boldsymbol{x}_0;t)\right]\nonumber\\
 &=& g^F_{\mu\nu}(\boldsymbol{x}_0;0).
\end{eqnarray} 
Eq.~(\ref{IUD}), i.e., Eq.~(\ref{LiouE}) is thus proved.
\end{proof}

\section{Proof of Eq.~(\ref{DCPTP}): The IDQS is not increased under parameter-independent  CPTP map.\label{contraction}}

Firstly, we prove that the IDQS is not increased under parameter-independent CPTP map $\Lambda_{t',t}$ which reads
\begin{equation}
 \mathcal{D}_Q(\boldsymbol{x}_0;t)\geqslant\mathcal{D}_Q(\boldsymbol{x}_0;t'),\label{ICPTP}
\end{equation}
with $\rho(\boldsymbol{x}_0;t')=\Lambda_{t',t}[\rho(\boldsymbol{x}_0;t)]$ and $t'\geqslant t$.
\begin{proof}
We begin with the coordinates $\boldsymbol{y}_0$ in which $\boldsymbol{g}^F(\boldsymbol{y}_0;t)$ is diagonal, i.e., $\boldsymbol{g}^F_{\mu\nu}(\boldsymbol{y}_0;t)=0$ for $\mu\neq \nu$. For the QFI is not increased under CPTP map $\Lambda_{t',t}$ in the single parameter cases, we have
\begin{equation}
\boldsymbol{g}^F_{\mu\mu}(\boldsymbol{y}_0;t)\geqslant\boldsymbol{g}^{F}_{\mu\mu}(\boldsymbol{y}_0;t'),\label{QFI-CPTP}
\end{equation} 
for all $\mu$.
For $g^{F}(\boldsymbol{y}_0;t')$ is positive semidefinite, we have
\begin{equation}
  \prod_\mu \boldsymbol{g}^{F}_{\mu\mu}(\boldsymbol{y}_0;t') \geqslant \left|\boldsymbol{g}^{F}(\boldsymbol{y}_0;t')\right|
\end{equation}
with the Hadamard's inequality.
Together with Eq.~(\ref{QFI-CPTP}), we have
\begin{equation}
 \left|\boldsymbol{g}^{F}(\boldsymbol{y}_0;t)\right|\geqslant \left|\boldsymbol{g}^{F}(\boldsymbol{y}_0;t')\right|.\label{DI}
\end{equation}
Multiplying $|\partial\boldsymbol{y}_0/\partial\boldsymbol{x}_0|$ on both sides of Eq.~(\ref{DI}), we have thus proved Eq.~(\ref{ICPTP}). 
\end{proof}
The equality in Eq.~(\ref{ICPTP}) is reached by the unitary channel. It is the direct results of Eq.~(\ref{IUD}). Furthermore, we have $\mathcal{D}_Q(\boldsymbol{x}_0;t)\geqslant0$ with the positive semi-definiteness of QFM. Eq.~(\ref{DCPTP}) is thus proved.

\section{\label{SLD}IDF with the symmetric logarithmic derivative}

Firstly, we expand the IDF as
\begin{equation}
 \mathcal{I}(\boldsymbol{x}_0;t)\equiv\frac{d}{d t}\left| \boldsymbol{g}^F\right|^\frac{1}{2}
 =\frac{1}{2}|\boldsymbol{g}^F|^\frac{1}{2}\mathrm{tr}\left[\frac{d}{d t}\log\boldsymbol{g}^F\right],\label{QFIF}
\end{equation}
with $\mathrm{tr}[\frac{d}{dt}\log\boldsymbol{g}^F]=\mathrm{tr}[\frac{d}{dt}\boldsymbol{g}^F/\boldsymbol{g}^F]$, $\boldsymbol{g}^F\equiv\boldsymbol{g}^F(\boldsymbol{x}_0;t)$ for simplicity, and the element 
\begin{equation}
 \frac{d}{dt} g^F_{\mu\nu}=\frac{1}{2}\mathrm{tr}[L_\nu \{\rho,\dot{L}_\mu \} +L_\mu  \{\rho,\dot{L}_\nu \} +\{ L_\mu , L_\nu \} \dot{\rho}],
\end{equation}
with $\dot{A}\equiv\frac{d}{dt}A$ for sccinctness.
For the derivative 
\begin{equation}
 \frac{d}{dt}\partial_\mu \rho=\frac{1}{2}\mathrm{tr}\left[\{\dot{\rho},L_\mu \}+\{\rho,\dot{L}_\mu \}\right]\label{eq:td},
\end{equation}
we have the anti-commutaor
\begin{align}
	\{\rho,\dot{L}_\mu \}
 & = 2\partial_\mu \dot{\rho}-\{\dot{\rho},L_\mu \} = (2\partial_\mu-L_\mu)\dot{\rho}-\dot{\rho}L_\mu.
\end{align}
Insert it into Eq.~(\ref{eq:td}), we have 
\begin{align}
 \frac{d}{dt} g^F_{\mu\nu}(\boldsymbol{x}_0;t)&=\mathrm{Tr}\left[\mathcal{L}_{\mu\nu}(\dot{\rho})\right],\label{PaSim} 
\end{align}
with the operator
\begin{align} 
 \mathcal{L}_{\mu\nu}=\frac{1}{2}\left[L_\nu(2\partial_\mu-L_\mu)+L_\mu (2\partial_\nu-L_\nu)\right].
\end{align} 

\section{\label{FTLME}Fisher information flow with time-local master equation}
In this appendix, we give the form of IDF under the time-local master equation Eq.~(\ref{TLME}).  We begin with the trace
\begin{widetext} 
\begin{align}
   & \mathrm{tr}\left[L_\nu \left(2\partial_\mu -L_\mu \right)\frac{d}{dt}\rho\right]_i\nonumber\\
= & \mathrm{tr}\left[L_\nu \left(2\partial_\mu -L_\mu \right)\left(A_i\rho A_i^{\dagger}-\frac{1}{2}\{A_i^{\dagger}A_i,\rho\}\right)\right]\nonumber\\
= & \mathrm{tr}\left[L_\nu A_i(L_\mu \rho+\rho L_\mu )A_i^{\dagger}-L_\nu L_\mu A_i\rho A_i^{\dagger}-\frac{1}{2}L_\nu \{A_i^{\dagger}A_i,L_\mu \rho+\rho L_\mu \}+\frac{1}{2}L_\nu L_\mu \{A_i^{\dagger}A_i,\rho\}\right]\nonumber\\
= & \mathrm{tr}\left[A_i^{\dagger}L_\nu A_iL_\mu \rho+L_\mu A_i^{\dagger}L_\nu A_i\rho-A_i^{\dagger}L_\nu L_\mu A_i\rho\right]\nonumber\\
   & \;-\frac{1}{2}\mathrm{tr}\left[L_\nu A_i^{\dagger}A_iL_\mu \rho + L_\mu L_\nu A_i^{\dagger}A_i\rho + A_i^{\dagger}A_iL_\nu L_\mu \rho + L_\mu A_i^{\dagger}A_iL_\nu \rho - L_\nu L_\mu A_i^{\dagger}A_i\rho - A_i^{\dagger}A_iL_\nu L_\mu \rho\right]\nonumber\\
 = & \mathrm{tr}\left[A_i^{\dagger}L_\nu A_iL_\mu \rho+L_\mu A_i^{\dagger} L_\nu A_i\rho-A_i^{\dagger}L_\nu L_\mu A_i\rho\right]\nonumber\\
   &\; -\frac{1}{2}\mathrm{tr}\left[L_\nu A_i^{\dagger}A_iL_\mu \rho + L_\mu A_i^{\dagger}A_iL_\nu \rho + L_\mu L_\nu A_i^{\dagger}A_i\rho - L_\nu L_\mu A_i^{\dagger}A_i\rho\right].
\end{align}
Insert it into Eq.~(\ref{PaSim}), we have the derivatives
\begin{align}
   & \frac{d}{dt}g^F_{\mu\nu}  (\boldsymbol{x}_0;t)_i\nonumber\\
 = & \frac{1}{2}\mathrm{tr}\left\{ \left[L_\nu \left(2\partial_\mu -L_\mu \right)+L_\mu \left(2\partial_\nu -L_\nu \right)\right]\frac{d}{dt}\rho\right\} _i\nonumber\\
 = & \frac{1}{2}\mathrm{tr}\left\{ A_i^{\dagger}L_\nu A_iL_\mu \rho+L_\mu A_i^{\dagger}L_\nu A_i\rho+A_i^{\dagger}L_\mu A_iL_\nu \rho+L_\nu A_i^{\dagger}L_\mu A_i\rho\right.\nonumber\\
   & \qquad\left.-L_\nu A_i^{\dagger}A_iL_\mu \rho-L_\mu A_i^{\dagger}A_iL_\nu \rho-A_i^{\dagger}L_\nu L_\mu A_i\rho-A_i^{\dagger}L_\mu L_\nu A_i\rho\right\}\nonumber\\
 = & -\frac{1}{2}\mathrm{tr}\left\{ \left([A_i,L_\nu ]^{\dagger}[A_i,L_\mu ]+[A_i,L_\mu ]^{\dagger}[A_i,L_\nu]\right)\rho\right\}. 
\end{align}
\end{widetext}
It indicates the derivative matrix $(d\boldsymbol{g}^F(\boldsymbol{x}_0,t)/dt)_i$ is negative semi-definite. We can diagonalize
it with a real orthonormal matrix $\boldsymbol{O}$ as 
\begin{equation}
 \boldsymbol{O}\left(\frac{d}{d t}\boldsymbol{g}^F\right)_i \boldsymbol{O}^T=\mathrm{diag}\left[\lambda_1^{(i)},\lambda_2^{(i)},\dots,\lambda_d^{(i)}\right],
\end{equation}
with the element
\begin{eqnarray}
\lambda_k ^{(i)}&= & \sum\nolimits_{\mu,\nu}O_{k\mu}\left(\frac{d}{d t}\boldsymbol{g}^F_{\mu\nu}\right)_i O_{\nu k}^T\nonumber\\
&= & -\mathrm{tr}\left\{ \left[A_i,\sum\nolimits_{\nu}O_{k\nu}L_{\nu}\right]^{\dagger}\left[A_i,\sum\nolimits_{\mu}O_{k\mu}L_{\mu}\right]\rho\right\}\nonumber\\
&\leqslant & 0.\label{DD}
\end{eqnarray}
Furthermore, $1/\boldsymbol{g}^F(\boldsymbol{x}_0;t)$ is positive definite, it indicates
\begin{equation}
 \alpha_\mu \equiv \left(\boldsymbol{O}\frac{1}{\boldsymbol{g}^F}\boldsymbol{O}^T\right)_{\mu\mu} >0.\label{Fd}
\end{equation}
Hence, we have the trace
\begin{equation}
 \mathrm{tr}\left[\frac{1}{\boldsymbol{g}^F}\left(\frac{d}{dt}\boldsymbol{g}^F\right)_i\right]=\sum_\mu \lambda_\mu^{(i)} \alpha_{\mu}\leqslant0.
\end{equation}
Insert it into Eq.~(\ref{Sub-IDF}), we have thus proved Eq.~(\ref{Subflow}) together with $\mathcal{D}(\boldsymbol{x}_0;t)\geqslant0$.

\end{document}